\documentclass{appolb}
\usepackage{graphicx}
\usepackage{cite}

\begin{document}
\title{Study of the $\eta$ meson production with polarized proton beam%
\thanks{Presented at II International Symposium on Mesic Nuclei}%
}
\author{M.~Hodana, P.~Moskal, I.~Ozerianska, M.~Zieli{\'n}ski
\address{The Marian Smoluchowski Institute of Physics, Jagiellonian University\\ Reymonta 4, Krak\'{o}w, Poland}\\
and\\
\address{Institut f\"{u}r Kernphysik, (IKP), Forschungszentrum J\"{u}lich\\ Wilhelm-Johnen-Stra{\ss}e 52428 J\"{u}lich, Germany} 
\\
and\\
\address{the WASA-at-COSY Collaboration}
}
\maketitle
\begin{abstract}
The $pp \to pp \eta$ reaction was investigated at excess energies of 15~MeV and 72~MeV
using the azimuthally symmetric WASA detector and a polarized proton beam of
the Cooler Synchrotron COSY.
The aim of the studies is the determination of partial wave
contributions to the production process of the $\eta$ meson in nucleon-nucleon collisions.
Here we present preliminary results of the extraction of the position of the interaction region
with respect to the WASA detector and preliminary results on the degree of polarization 
of the COSY proton beam used in the experiment.
\end{abstract}
  
\section{Introduction}

In spite of the number of both experimental~\cite{chiavassa94,calen96,calen97,hibou98,smyrski00,bergdolt93,abdelbary03,
ETA-PRC-Moskal,ETA-EPJ-Moskal,ETA-Petren,pn-deta-Calen,pn-pneta-Calen,pn-pneta-Moskal} 
and theoretical~\cite{ACTA-Colin,nakayama02,faldt01,germond90,laget91,moalem96,vetter91,alvaredo94,batinic97} studies performed 
so far for measurements of total and differential cross sections for the $\eta$ meson production in nucleon-nucleon collisions, 
the proton-$\eta$ interaction as well as the mechanism of the $\eta$ meson production have not been fully elucidated yet.
From the above cited measurements of the $\eta$ meson production in $pp$ and $pn$ reactions we learned that 
the production occurs predominantly via the $N(1535)$ resonance and that the proton-$\eta$ interaction is much 
larger than in case of proton-$\pi^0$, and proton-$\eta'$ interactions~\cite{moskal00,review}.
The knowledge of the $\eta$ and $\eta^{\prime}$ meson interaction with nucleons is crucial for the search of the
mesic nuclei which is recently carried out in many laboratories, e.g. 
  COSY~\cite{COSY11-MoskalSmyrski,WASA-at-COSY-SkuMosKrze,Adlarson2013,GEM},
  ELSA~\cite{cbelsa},
  GSI~\cite{eta-prime-mesic-GSI-Itahashi},
  JINR~\cite{JINR},
  JPARC~\cite{eta-mesic-JPARC-Fujioka},
  LPI~\cite{LPI}, and
  MAMI~\cite{ELSA-MAMI-plan-Krusche} with the increasing
theoretical support e.g.~\cite{wilkin2,bass,eta-prime-mesic-Nagahiro,Mesic-Bass,EtaMesic-Hirenzaki,
ETA-Friedman-Gal,Wycech-Acta,Mesic-Kelkar,Bass10Acta,uzikov,niskanen,sibirtsev}.
Previous studies of the $\eta$ meson production in collisions of nucleons revealed that even 
in the close-to-threshold region higher partial waves and other baryon resonances may contribute  
to the production mechanism. Moreover, the indication of the contribution of higher partial waves near threshold 
comes also from the comparison
of the invariant mass distribution from the production of pp$\eta$ and pp$\eta^{\prime}$ systems~\cite{ETAPRIME-Klaja}.
Therefore, for an unambiguous understanding of the production process relative magnitudes 
from the partial wave contributions must be well established. 
This may be at least to some extent achieved by the measurement of the analyzing 
power $A_{y}$ which would enable to perform the partial wave
decomposition with an accuracy by far better than 
resulting from the measurements of the distributions of the spin averaged 
cross sections.
Up to now, measurements of the analyzing power for the $\vec{p}p\to pp\eta$ reaction were performed
by the COSY-11 and DISTO collaborations~\cite{aycosy11,ETA-Ay-EPJ-Winter,ETA-Ay-PL-Winter,ETA-Ay-Balestra}. 
Due to the lack of statistics and small detector acceptance (in case of COSY-11~\cite{Brauksiepe,C11Klaja})
these first measurements did not allow for unambiguous conclusions about the production mechanisms.
Therefore, a high statistics measurement was made with the large acceptance ($\sim 4\pi$) symmetric WASA detector~\cite{HHAdam}. 
The experiment was conducted for beam momenta of 2026 MeV/c and 2188 MeV/c\cite{spin2010} which 
correspond to excess energies of 15~MeV and 72~MeV, respectively.
To monitor the degree of polarization, the luminosity and the detector performance,
simultaneously the $\vec{p}p\to pp$ reaction was measured.
In order to control effects caused by the potential asymmetries in the detector setup, 
the spin direction of the proton beam was flipped from cycle to cycle.

In the next sections we briefly describe the experiment and remind the conclusions drawn from simulations studies performed 
so far \cite{Hodana:2013cga} and after that we present preliminary results from the studies of the degree of polarization 
of the proton beam used in the experiment.

\section{Studies of $A_{y}$ with the WASA-at-COSY detector}
The axially symmetric WASA detector and the vertically polarized proton beam of
COSY have been used to collect a high statistics sample of $\vec{p}p \rightarrow pp\eta$ reactions in order to determine the
analyzing power as a function of the invariant mass spectra of the two particle subsystems and as a function 
of the emission angle of the $\eta$ meson~\cite{Prop10}. 

For the monitoring of degree of polarization, simultaneously to the $\vec{p}p \rightarrow pp\eta$ 
reaction, the proton-proton elastic scattering reaction has been measured. The estimation  of systematic 
uncertainties of the determination of the degree of polarization of the 
beam are presented in~\cite{Hodana:2013cga}. 
Performed analyses revealed that to reach a systematic uncertainty of the polarization smaller than
$3\%$, the position of the center of the interaction region has to be controlled with a precision better than $1$~mm.
The large statistics of collected data and utilization of methods of vertex reconstruction shown
in~\cite{Hodana:2013cga,Demirors2005}, allowed us to determine the average vertex position with the precision much better than $1$~mm. 
Furthermore, conducted studies show that the beam tilted within the maximum allowed range should
have no significant influence on the obtained degree of polarization~\cite{Hodana:2013cga}.

\subsection{Extraction of the average vertex positions from the experimental data}
\begin{figure}
\includegraphics[width=0.47\linewidth]{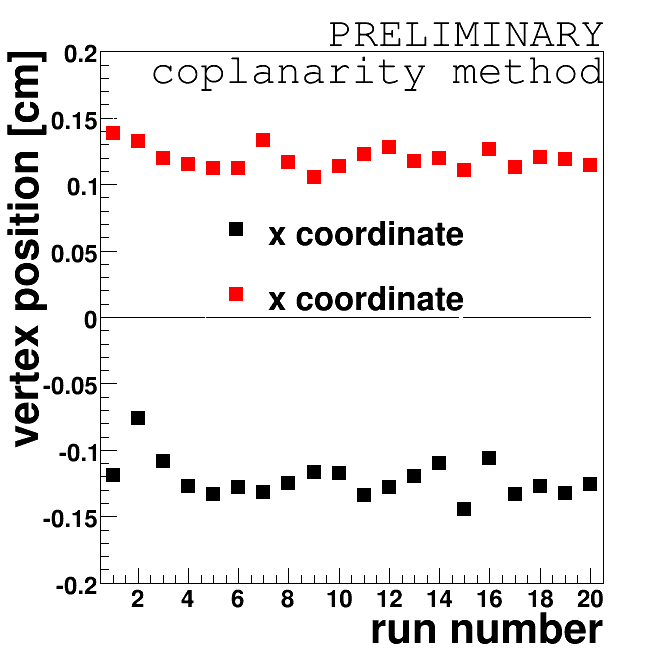}
\includegraphics[width=0.47\linewidth]{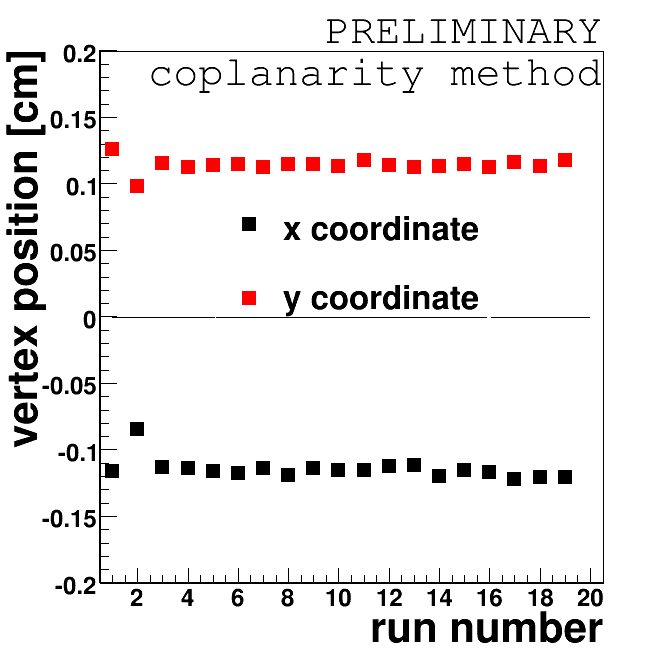}\\
\includegraphics[width=0.47\linewidth]{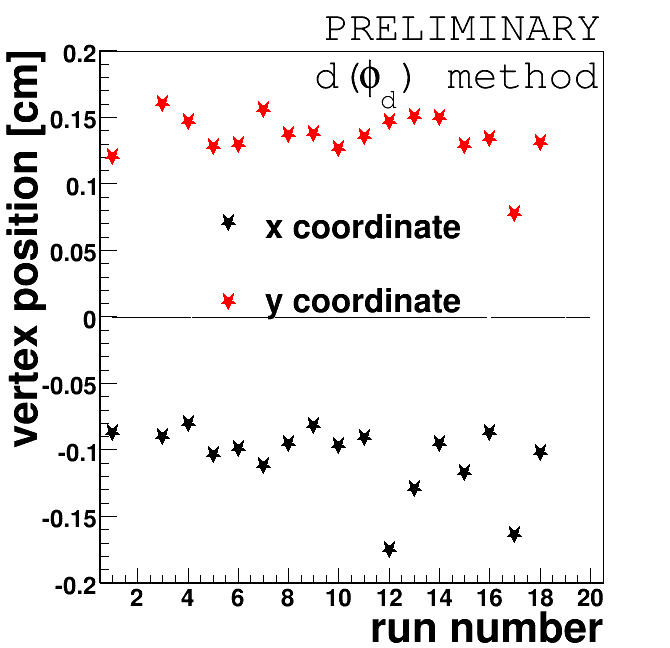}
\includegraphics[width=0.47\linewidth]{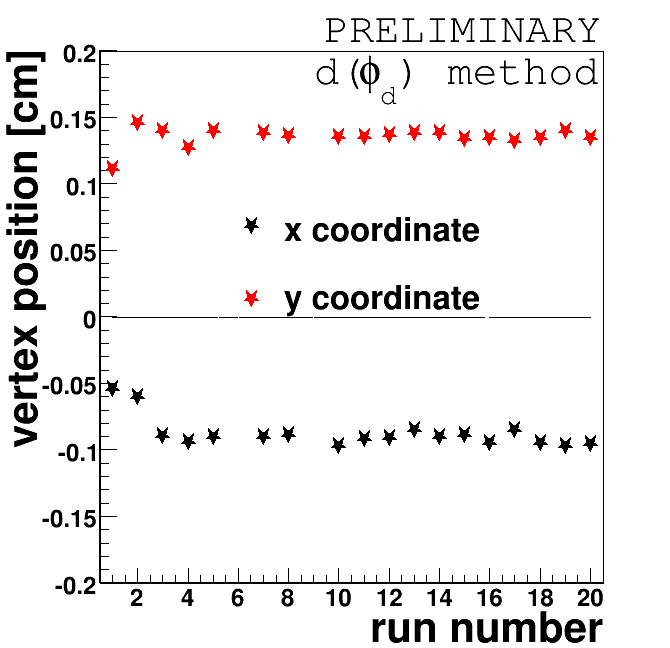}

\caption{Distributions of the shift from the nominal value of a given coordinate of the center of the interaction 
region as a function of 
time (run number). Plots were made for data collected with an unpolarized beam (left column)
and with a polarized beam (right column). The results obtained using the coplanarity-method are 
shown in the upper row. In the lower row the results obtained using the $d(\phi_{d})$-method 
\cite{Hodana:2013cga} are presented.}
\label{f:PosRun}
\end{figure}

To find the position of the vertex, $(v_{x},v_{y},v_{z})$ in the experiment, methods described in \cite{Demirors2005,Hodana:2013cga}
have been applied. The first utilized method is based on the angular dependence of the coplanarity of incoming and outgoing protons,
which is defined as:
\begin{equation}
C = \frac{( \vec{p}_{1}\times \vec{p}_{2}) \cdot \vec{p}_{beam}}{ |\vec{p}_{1}\times \vec{p}_{2}| \cdot |\vec{p}_{beam}|},
\end{equation}
where $\vec{p}_{1}$ and $\vec{p}_{2}$ corresponds to momentum vectors of scattered protons, and $\vec{p}_{beam}$ is the  
beam momentum vector. In order to find the center of the interaction region, coplanarity distributions as a function 
of $\phi$ angle simulated with different vertex positions are compared with the experimental one using the $\chi^{2}$ statistics. 
For each $C(\phi)$ spectrum a $\chi^{2}$ value is calculated according to:
\begin{equation}
\chi^2 = \sum_{i}\frac{(M_{i}^{MC} - M_{i}^{exp})^{2}}{(\sigma_{i}^{exp})^2},
\end{equation} 
where $i$ indicates the chosen $\phi$ range, the $M_{i}^{MC}$ and $M_{i}^{exp}$ are the mean values 
of the cooplanarities in a given $\phi$ range and $\sigma_{i}^{exp}$ is the error of $M_{i}^{exp}$.
The corresponding distributions of the vertex shift for a given coordinate as a function of time 
(for twenty exemplary runs) are shown in Fig.~\ref{f:PosRun}. Analyses were performed for both
data sets: with polarized beam (upper left) and unpolarized beam (upper right). 




The second method is based on utilization of the $d(\phi_{d})$ distributions 
as shown in \cite{Hodana:2013cga}. The resulting experimental spectra of position of a given coordinate as function of 
time (run number) are shown in Fig.~\ref{f:PosRun} in the lower row (left and right). One can see that
for the data with polarized beam, the vertex position is relatively stable with time, however for the data sample collected with unpolarized  beam some fluctuations are observed. 
Nevertheless both methods give results for $v_{x}$ and $v_{y}$ coordinates that differs on the average
only by about $0.04$~cm.
Thus we may conclude that at the present stage of experimental data
analysis the systematic uncertainty of the determination of the position of the interaction region 
is equal to about $\pm$~0.2~mm which correspond to an uncertainty of the polarization determination of less than 
$\pm~1\%$~(see figures in~\cite{Hodana:2013cga}).

%


\subsection{Extraction of the degree of polarization from experimental data}

The method of polarization determination is described in detail in~\cite{Hodana:2013cga}.
Therefore, for the sake of completeness we only briefly recall that the polarization P is extracted 
by fitting the experimental distributions with the function~\cite{Hodana:2013cga}:
\begin{equation}
\epsilon(\theta,\phi) = P(\theta) \cdot A_{y}(\theta) \cdot cos( \phi ),
\end{equation}
\begin{figure}
\includegraphics[width=\linewidth]{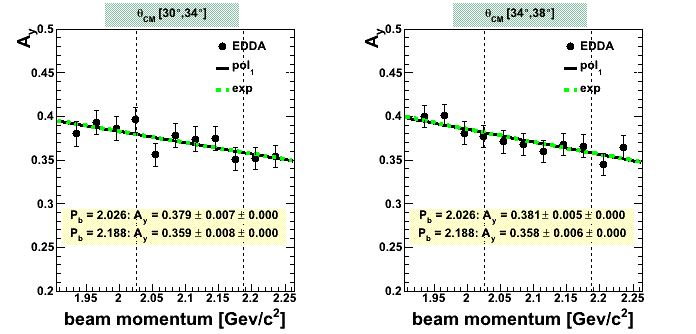}
\caption{The $A_{y}(\theta_{CM},p_{beam})$ distributions obtained by the EDDA collaboration. Data points are
					shown as filled circles. Fitted functions are described in the legend. Dashed horizontal lines 
					mark the two beam momenta for which WASA data were taken. For both beam momenta, evaluated analyzing 
					powers are shown with the statistical and systematic errors respectively.}
\label{f:edda}
\end{figure}
%
  
where the asymmetry
\begin{equation}
\epsilon(\theta,\phi) =\frac{N(\theta,\phi) - N(\theta,\phi+\pi)}{N(\theta,\phi) + N(\theta,\phi+\pi)}
\label{e:Asym}
\end{equation}
is calculated separately for each spin orientation of polarized protons, 
in two ranges of proton scattering angles of $30^\circ-34^\circ$ and $34^\circ-38^\circ$.
To obtain $A_{y}$ at a desired beam momentum and to estimate a systematic uncertainty of this determination, two different functions 
are fitted to the momentum dependence of $A_{y}$ measured by the EDDA collaboration~\cite{Altmeier2000}
in these angular ranges. The plots used for the extraction are shown in Fig.~\ref{f:edda}. 
As a result, two polarizations are extracted for two ranges of the center-of-mass polar angle of the forward scattered proton,
and a weighted mean is used as a final polarization for a given spin orientation~\cite{Hodana:2013cga}.

\begin{figure}
\includegraphics[width=0.47\linewidth]{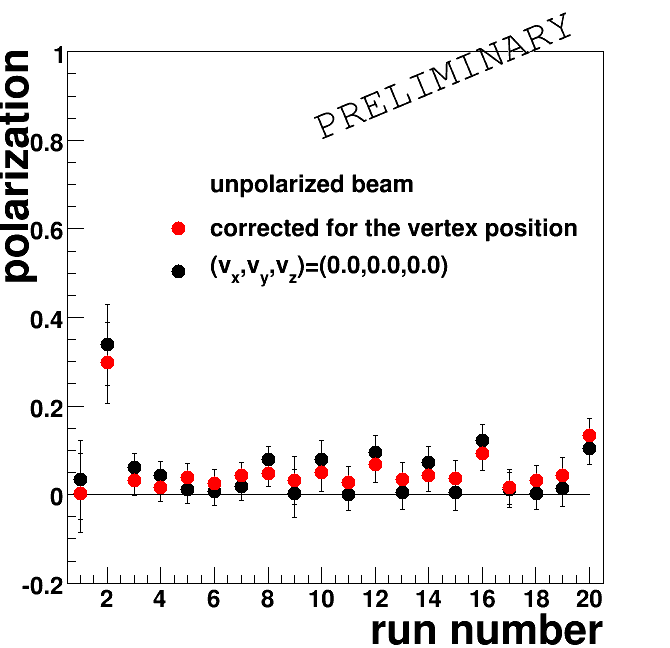}
\includegraphics[width=0.47\linewidth]{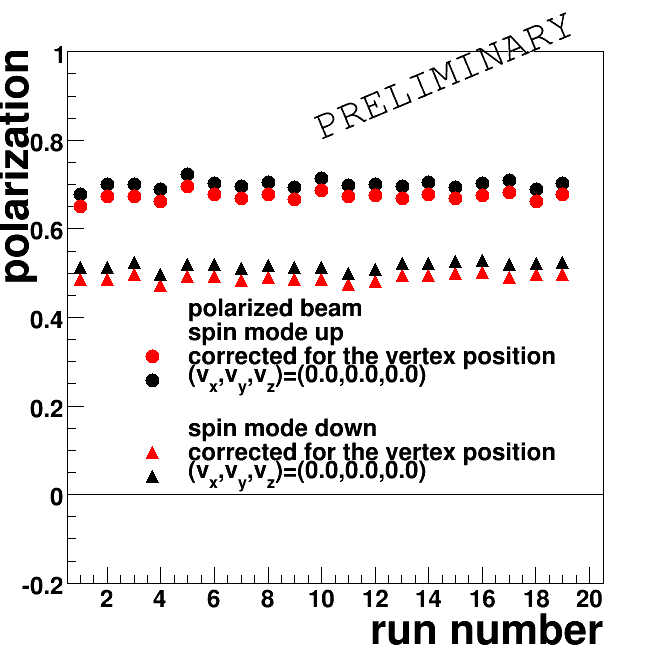}
\caption{Distributions of polarization as a function of run number for unpolarized (left) and polarized (right) data
(taken at a beam momentum of $p_{beam}=2026$~MeV/$c$).
Data points have been acceptance corrected using the default vertex position at $(v_{x},v_{y},v_{z}) = (0,0,0)$ (black
marker) and the vertex position established based on the experimental data (red marker). Results for both polarization
modes of the beam particles are shown.}
\label{f:Pol}
\end{figure}

The polarization for twenty runs (about $5\%$ of data) is shown in Fig.~\ref{f:Pol}. On the left panel, the polarization obtained from
data collected with an unpolarized beam is presented and, therefore, should be consistent with zero.
On the right panel, the results obtained from the analysis of data gathered with polarized beam are shown.
The polarization was calculated for both orientations of proton spin separately. Data points shown in Fig.~\ref{f:Pol}
have been corrected for acceptance determined using the vertex position extracted from the experimental data. For 
comparison, also the result assuming a nominal center of the vertex region 
$(v_{x},v_{y},v_{z}) = (0,0,0)$ is plotted.

\section{Summary}
Preliminary results of the extraction of the  $v_{x}$ and $v_{y}$ coordinates 
of the center of the interaction region have been shown. 
At the present stage of analysis the systematic uncertainty in the determination of the position of the interaction region 
is equal to about $\pm$~0.2~mm which corresponds to an uncertainty of the polarization determination of less than 
$\pm 1\%$~(see figures in~\cite{Hodana:2013cga}).
The polarization for the measurement with a beam momentum of $p_{beam}=2026$~MeV/c
was determined preliminary to be about 49\% and 67\% for spin down and spin up orientations, respectively.
For the measurement with unpolarized beam a small but non-zero value of polarization (4\%) was found even after the correction 
for the average position of the interaction points. Therefore, further detailed studies of the possible 
reason of the non-zero polarization for the unpolarized beam are required.  
However, it should be stressed that in this contribution we show that the collected data are of a high quality with 
average polarization of about 58\%. It was also shown that it should be possible to control the degree of polarization
with a systematic precision of about $\pm 1\%$. 

\subsection{Acknowledgements}
We acknowledge support 
by the Polish National Science Center through grant No. 2011/03/B/ST2/01847, 
by the FFE grants of the Research Center J\"{u}lich, 
by the EU Integrated Infrastructure Initiative HadronPhysics Project under contract number RII3-CT-2004-506078 
by the European Commission under the 7th Framework Programme through the ’Research Infrastructures’ action of the ’Capacities’ Programme, Call: FP7~-~INFRASTRUCTURES~-~2008~-~1, Grant Agreement N. 227431, 
and by the Polish Ministry of Science and Higher Education through grant No. 393/E-338/STYP/8/2013.


\begin{thebibliography}{9}
\bibitem{chiavassa94} E. Chiavassa et al., ~\textit{Phys. Lett.} \textbf {B322} (1994) 270
\bibitem{calen96} H. Cal\'{e}n et al., ~\textit{Phys. Lett.} \textbf {B366} (1996) 39
\bibitem{calen97} H. Cal\'{e}n et al., ~\textit{ Phys. Rev. Lett.} \textbf {79} (1997) 2642
\bibitem{hibou98} F. Hibou et al., ~\textit{Phys. Lett.} \textbf {B438} (1998) 41
\bibitem{smyrski00} J. Smyrski et al., ~\textit{Phys. Lett.} \textbf {B474} (2000) 182
\bibitem{bergdolt93} A.M. Bergdolt et al., ~\textit{Phys. Rev.} \textbf {D48} (1993) 2969
\bibitem{abdelbary03} M. Abdel-Bary et al., ~\textit{Eur. Phys. J.} \textbf {A16} (2003) 127
\bibitem{pn-pneta-Calen} H. Cal\'{e}n et al., ~\textit{Phys. Rev.} \textbf {C 58} (1998) 2667 
%
\bibitem{ETA-PRC-Moskal} P. Moskal et al.,
     ~\textit{Phys. Rev.} \textbf {C69} (2004) 025203
%
\bibitem{ETA-EPJ-Moskal} P. Moskal et al.,
     ~\textit{Eur. Phys. J.} \textbf {A43} (2010) 131
%
\bibitem{ETA-Petren} H. Petren~et al.,
     ~\textit{Phys. Rev.} \textbf {C82} (2010) 055206
%
%
\bibitem{pn-deta-Calen} H. Calen et al.,
     ~\textit{Phys. Rev. Lett.} \textbf {79} (1997) 2642
%
\bibitem{pn-pneta-Moskal} P. Moskal et al.,
     ~\textit{Phys. Rev.} \textbf {C79} (2009) 015208
%
\bibitem{ACTA-Colin} C. Wilkin,
         ~\textit{Acta Phys. Polon.} \textbf {B41} (2010) 2191

\bibitem{nakayama02} K. Nakayama et al., ~\textit{Phys. Rev.} \textbf {C65} (2002) 045210
\bibitem{faldt01} G. F\"{a}ldt and C. Wilkin, ~\textit{Phys. Scripta} \textbf {64} (2001) 427
\bibitem{germond90} J. F. Germond et al., ~\textit{Nucl. Phys.} \textbf {A518} (1990) 308
\bibitem{laget91} J. M. Laget et al., ~\textit{Phys. Lett.} \textbf {B257} (1991) 254
\bibitem{moalem96} A. Moalem et al., ~\textit{Nucl. Phys.} \textbf {A600} (1996) 445
\bibitem{vetter91} T. Vetter et al., ~\textit{Phys. Lett.} \textbf {B263} (1991) 153
\bibitem{alvaredo94} B. L. Alvaredo et al., ~\textit{Phys. Lett.} \textbf {B324} (1994) 125
\bibitem{batinic97} M. Batini\,{c} et al., ~\textit{Phys. Scripta} \textbf {56} (1997) 321
\bibitem{moskal00} P. Moskal et al., ~\textit{Phys. Lett.} \textbf {B482} (2000) 356
\bibitem{review} P. Moskal et al., ~\textit{Prog. Part. Nucl. Phys.} \textbf {49} (2002) 1
%
\bibitem{COSY11-MoskalSmyrski}
   P. Moskal, J. Smyrski, ~\textit{Acta Phys. Pol.} \textbf {B41} (2010) 2281 
%
\bibitem{WASA-at-COSY-SkuMosKrze}
  M. Skurzok, P. Moskal, W. Krzemien, ~\textit{Prog. Part. Nucl. Phys.} \textbf {67} (2012) 445 
%
\bibitem{Adlarson2013}
    P. Adlarson et al., ~\textit{Phys. Rev.} \textbf{C87} (2013) 035204
\bibitem{GEM}
A. Budzanowski et al., ~\textit{Phys. Rev.} \textbf{C79} (2009) 012201
\bibitem{cbelsa} M.Nanova al., ~\textit{Phys. Lett.} \textbf{B727} (2013) 417 
\bibitem{eta-prime-mesic-GSI-Itahashi} K. Itahashi al., ~\textit{Prog. Theor. Phys.} \textbf {128} (2012) 601 
%
\bibitem{JINR} S.V. Afanasiev,
      ~\textit{Phys. Part. Nucl. Lett.} 8 (2011) 1073
\bibitem{eta-mesic-JPARC-Fujioka} H. Fujioka, ~\textit{Acta Phys. Polon.} \textbf {B41} (2010) 2261 
%
\bibitem{LPI}V.A. Baskov  et al.,
      ~\textit{PoS Baldin-ISHEPP-XXI} (2012) 102, arXiv:1212.6313 [nucl-ex] 
\bibitem{ELSA-MAMI-plan-Krusche} B. Krusche et al., ~\textit{J. Phys. Conf. Ser.}, \textbf{349} (2012) 012003 
\bibitem{wilkin2} C.~Wilkin, ~\textit{Phys. Lett.} \textbf {B654} (2007) 92 
\bibitem{bass} S.D. Bass, A.W. Thomas, ~\textit{Phys. Lett.} \textbf{B634} (2006) 368 
%
\bibitem{eta-prime-mesic-Nagahiro} Nagahiro H.~\textit{et al.}:
        ~\textit{Phys. Rev.} \textit{C87} (2013) 045201
%
\bibitem{Mesic-Bass} S.D. Bass, A.W. Thomas, e-Print: arXiv:1311.7248[hep-ph], ~\textit{Acta Phys. Pol.} \textbf{B45} (2014), in print.
%
\bibitem{EtaMesic-Hirenzaki} S. Hirenzaki et al., ~\textit{Acta Phys. Polon.} \textbf{B41} (2010) 2211 
\bibitem{ETA-Friedman-Gal} E. Friedman, A. Gal, J. Mares,
   ~\textit{Phys. Lett.} \textbf {B725} (2013) 334
\bibitem{Wycech-Acta} S. Wycech, W. Krzemien, e-Print: arXiv:1401.0747 [nucl-th], \textit {Acta Phys. Pol.} \textbf {B45} (2014)
%
%
\bibitem{Mesic-Kelkar} N. G. Kelkar et al., ~\textit{Rept. Prog. Phys.} \textbf {76} (2013) 066301
%
\bibitem{Bass10Acta} S. D. Bass, A. W. Thomas, e-Print: arXiv:1007.0629 [hep-ph], ~\textit{Acta Phys.Polon.} \textbf {B41} (2010) 2239
\bibitem{uzikov} Yu.N. Uzikov, ~\textit{Nucl. Phys.} \textbf {A801} (2008) 114
\bibitem{niskanen} J.A. Niskanen, arXive:1312.7281
\bibitem{sibirtsev} A. Sibirtsev et al., ~\textit{Phys. Rev.} \textbf {C70} (2004) 047001
\bibitem{ETAPRIME-Klaja} P. Klaja et al.,
    ~\textit{Phys. Lett.} \textbf {B684} (2010) 11
%
\bibitem{aycosy11} R.~Czy{\.z}ykiewicz et al., ~\textit{Phys. Rev. Lett.} \textbf {98} (2007) 122003
\bibitem{ETA-Ay-EPJ-Winter} P. Winter et al., ~\textit{Eur. Phys. J.} \textbf {A18} (2003) 355
\bibitem{ETA-Ay-PL-Winter} P. Winter et al., ~\textit{Phys. Lett.} \textbf {B544} (2002) 251; Erratum-ibid. B553 (2003) 339
\bibitem{ETA-Ay-Balestra} F. Balestra et al., ~\textit{Phys. Rev.} \textbf {C69} (2004) 064003
\bibitem{C11Klaja}
P. Klaja et al., ~\textit{AIP Conf. Proc.} \textbf {796} (2005) 160
\bibitem{Brauksiepe} S.~Brauksiepe et al.,~\textit{ Nucl. Instrum. Methods Phys. Res.} \textbf{A376} (1996) 397    
\bibitem{spin2010} P. Moskal, M. Hodana,~\textit{ J. Phys. Conf. Ser}. 295 (2011) 012080;  e-Print: arXiv:1101.5486
\bibitem{Hodana:2013cga} M. Hodana, P. Moskal, I. Ozerianska, ~\textit{Acta Phys. Polon. Supp.} \textbf{6} (2013) 1041
\bibitem{HHAdam} H. H. Adam et al., nucl-ex/0411038
\bibitem{Prop10} P. Moskal, M. Hodana and H.Cal\'{e}n, ~\textit{Proposal to COSY-PAC} \textbf {185.1} (2010)
\bibitem{Demirors2005} L. Demirors PhD Hamburg Univerity (2005)
\bibitem{Altmeier2000} M. Altmeier et al. ~\textit{Phys. Rev. Lett.} \textbf{85} (2000) 1819
\end{thebibliography}
\end{document}